\documentclass[11pt,a4paper]{article}
\usepackage{jcappub}

\usepackage{graphicx}
\usepackage{dcolumn}
\usepackage{bm}

\newcommand{\be}{\begin{equation}}
\newcommand{\en}{\end{equation}}
\newcommand{\bea}{\begin{eqnarray}}
\newcommand{\ena}{\end{eqnarray}}

\title{Semi-holographic model revisited}

\author{V\'ictor H. C\'ardenas}
\author{J. R. Villanueva}
\author{Juan Maga\~na}%

\affiliation{ Departamento de F\'{\i}sica y Astronom\'ia, Facultad
de Ciencias, Universidad de Valpara\'iso, Gran Breta\~na 1111,
Valpara\'iso, Chile,}
\affiliation{Centro de Astrof\'isica de Valpara\'iso, Gran Breta\~na 1111, Playa Ancha,
\\ Valpara\'{\i}so, Chile.}

\emailAdd{victor.cardenas@uv.cl}
\emailAdd{jose.villanuevalob@uv.cl}
\emailAdd{juan.magana@uv.cl}

\date{\today}

\abstract{In a recent work Zhang, Li and Noh [Phys. Lett. B {\bf 694}, 177 (2010)]
proposed a model
for dark energy assuming this component strictly obeys the
holographic principle. They performed a dynamical system analysis,
finding a scaling solution which is helpful to solve the coincidence
problem. However they need explicitly a cosmological constant. In
this paper we derive an explicit analytical solution, without
$\Lambda$, that shows agreement with the Supernovae data. However
this solution is not physical because violate all the energy
conditions.}


\begin{document}

\maketitle

\section{Introduction}

The accelerated expansion of the Universe is one of the biggest
challenges for science today. The source of this mysterious cosmic
acceleration is dubbed dark energy (DE). Although the first evidence
arrived from studies of Type Ia supernova (SNIa) \cite{01}, we have
strong indications also from large scale structure (LSS)\cite{02},
cosmic microwave background (CMB)\cite{03}, the integrated
Sachs--Wolfe effect (ISW)\cite{isw}, baryonic acoustic oscillations
(BAO)\cite{Eisenstein:2005su} and gravitational lensing
\cite{weakl}. The simplest way to fit these multiple data is by
introducing a constant $\Lambda$ in the theory; the so far
successful $\Lambda$ cold dark matter ($\Lambda$CDM) model. However
it suffers from two main problems: the low value of the vacuum
energy and the coincidence problem \cite{1}. Thus, it seems
necessary to assume the existence of a dynamical cosmological
constant, which leads to consider dynamical DE models as scalar
field models (e.g., quintessence \cite{2}, K-essence \cite{3} and
tachyon fields \cite{4}). Another way to consider dynamical models
is assume the existence of interactions in the dark sector (DE and
dark matter) \cite{interaction}. Because both dark components are
characterized through their gravitational effects, it is natural to
consider unified models of the cosmological substratum in which one
single component plays the role of DM and DE simultaneously.
Examples of this type of models are the Chaplygin gas \cite{5}, and
bulk-viscous models \cite{bv}.

The holographic dark energy is one of the emergent dynamical DE
model proposed in the context of fundamental principle of quantum
gravity, so called holographic principle. This principle arose due
to development in the study of black hole theory and string theory.
The holographic principle states that the number of degrees of
freedom of a physical system, apart of being constrained by an
infrared cut-off, it should be finite and it should scale with its
bounding area rather than with its volume\cite{7}. Specifically, it
is derived with the help of entropy-area relation of thermodynamics
of black hole horizons in general relativity which is also known as
the Bekenstein-Hawking entropy bound, i.e., $S \simeq M_p^2 L^2$,
where $S$ is the maximum entropy of the system of length $L$ and
$M_p=1/\sqrt{8\pi\,G}$ is the reduced Planck mass. In general terms,
the inclusion of the holographic principle into cosmology, it was
possible to find the upper bound of the entropy contained in the
universe\cite{FS98}. Using this idea Cohen \cite{cohen} suggested a
relation between the short distance (ultraviolet, UV) cutoff and the
long distance (infrared, IR) cutoff which, after identifying
infrared with the Hubble radius $H^{-1}$, resulted in a DE density
very close to the observed critical energy density. After this, Li
\cite{li} studied the use of both the particle and event horizons as
the IR cutoff length. He found that apparently only a future event
horizon cutoff can give a viable DE model. More recently, it was
proposed a new cutoff scale, given by the Ricci scalar curvature
\cite{021,022}, resulting in the so-called holographic Ricci DE
models.

In \cite{Zhang:2010iz} a new holographic model was proposed assuming
that DE obeys strictly an holographic principle. They performed a
dynamical system analysis finding a stable DM-DE solution which
ameliorate the coincidence problem, however they mentioned the
necessity to add an explicit cosmological constant to obtain an
accelerated expansion. Actually this claim was reinforced again in
\cite{Li:2012vw} where the authors proposed a generalization of the
original model.

In this paper we revisit the original model, and show that it is
possible to find an analytic solution that shows an accelerated
expansion evolution, and also we show that this solution is in
agreement with the latest supernovae data. However we also show that
this solution is not physical because it violates the
weak-energy-condition.

\section{The semi-holographic model}

In an adiabatically evolving universe, the first law of
thermodynamics equals the continuity equation. In a comoving volume
the first law reads, \be dU=TdS-pdV, \label{1st} \en where
$U=\frac{4}{3}\pi\rho a^3$ is the energy in this volume for the
spatially flat case, $T$ denotes temperature, $S$ represents the
entropy of this volume, and $V$ stands for the physical volume
$V=\frac{4}{3}\pi a^3$, $\rho$ is the energy density and $a$ denotes
the scale factor. In this paper we only consider the flat case.

In \cite{Zhang:2010iz} the authors write the entropy in the apparent
horizon in a comoving volume as
   \be
   S_c=\frac{8\pi^2 \mu^2}{H^2} \frac{a^3}{H^{-3}}={8\pi^2
   \mu^2}Ha^3.
   \label{Sc}
   \en
following \cite{bak, cai}. Here $H$ is the Hubble parameter, and
$\mu$ denotes the reduced Planck mass. The semi holographic model
emerge from here assuming the dark energy satisfy a similar relation
    \be
    S_{de}={8\pi^2
   \mu^2}Ha^3.
   \label{detro}
    \en
Because the expansion is adiabatic, the dark matter entropy
contribution is
    \be
    S_{dm}=C-S_{de},
    \label{dmtro}
    \en
where $C$ is a constant representing the total entropy of the
comoving volume. From the Friedmann equation we can obtain the
expressions to relate the entropies with the corresponding energy
densities, as
 \be
   H^2=\frac{1}{3\mu^2}(\rho_{dm}+\rho_{de}+\Lambda),
   \label{fried}
   \en
where $\rho_{dm}$ denotes the density of dark matter, $\rho_{de}$
denotes the density of dark energy. $\Lambda$ is a cosmological
constant term, which was claimed necessary in the original treatment
\cite{Zhang:2010iz}, however we can safely set to zero as we will
see in short. The holographic principle requires also that the
temperature \cite{cai} is related to the Hubble function as
   \be
   T=\frac{H}{2\pi}.
   \label{tem}
   \en
which close the necessary system of equations.

The dark energy evolution equation can be obtained from the
combination of (\ref{tem}), (\ref{fried}), and (\ref{detro}), in the
first law (\ref{1st}):
   \be
   \frac{2}{3}\rho_{de} '=\rho_{dm}(1-\omega_{dm})-\rho_{de}(1+3\omega_{de}),
   \label{evlde}
   \en
where a prime means $'=d/d\log a$, $\omega_{dm}$ indicates the
equation of state (EOS) parameter of dark matter, and $\omega_{de}$
represents the EOS parameter of dark energy. Using instead of
(\ref{detro}) the complementary relation (\ref{dmtro}) for dark
matter we find
\be
   \frac{2}{3}\rho_{dm} '=-\rho_{dm}(3+\omega_{dm})+\rho_{de}(-1+\omega_{de}).
   \label{evldm}
   \en

%
%

\section{The analytical solution}

Let us to consider the system (\ref{evlde}) and (\ref{evldm}). This
system can be written as ${\bf u'}=M{\bf u}$ where ${\bf
u}=(\rho_{de},\rho_{dm})$ is the vector containing the energy
densities and $M$ is the following matrix
\begin{equation}\label{matrix}
M = \frac{3}{2}\left[
      \begin{array}{cc}
        -1-3\omega_{de} & 1-\omega_{dm} \\
        -1+\omega_{de} & -3-\omega_{dm} \\
      \end{array}
    \right].
\end{equation}
A standard procedure leads to the analytic solution. They are linear
combinations of the roots $r$ of the quadratic equation
\begin{equation}\label{detM}
\det{( M-rI)}=0,
\end{equation}
Explicitly we found,
\begin{eqnarray}\label{as1}
\nonumber \rho_{de}= f(z)\left\{
a[1-(1+z)^{\frac{3\kappa}{2}}]+\rho_{de}^{(0)}[1+(1+z)^{\frac{3\kappa}{2}}]\right\}\\
\rho_{dm}= f(z)\left\{
b[1-(1+z)^{\frac{3\kappa}{2}}]+\rho_{dm}^{(0)}[1+(1+z)^{\frac{3\kappa}{2}}]\right\}
\end{eqnarray}
where
\begin{eqnarray}
\nonumber 
  a&=& \rho_{de}^{(0)}(2-3\omega_{de}+\omega_{dm})+ 2\rho_{dm}^{(0)}(1-\omega_{dm}), \\
\nonumber
  b &=& 2\rho_{de}^{(0)}(-1+\omega_{de})+\rho_{dm}^{(0)}(-2+3\omega_{de}-\omega_{dm}), \\
\nonumber
  \kappa &=&
  \sqrt{(\omega_{de}-\omega_{dm})(-8+9\omega_{de}-\omega_{dm})},
\end{eqnarray}
and finally the function
\begin{equation}\label{fdez}
f(z)=
\frac{1}{2\kappa}(1+z)^{\frac{3}{4}(4+3\omega_{de}+\omega_{dm}-\kappa)}.
\end{equation}
Clearly to get real energy densities first we need to impose the
condition
\begin{equation}\label{cond1}
(\omega_{de}-\omega_{dm})(-8+9\omega_{de}-\omega_{dm})\geq 0.
\end{equation}
We also need to consider the stability of the solution. The critical
points of the system are
\begin{equation}\label{as4}
  \rho_{de}^{(C)}=0,\qquad \rho_{dm}^{(C)}=0.
\end{equation}
Note, from (\ref{as1}) and (\ref{fdez}) that these critical points
occur at $z = -1$. However, as we shall see in section 5, depending on the initial conditions, we find that 
one of these solutions becomes zero for $ z> 0$.

In order to obtain a stability analysis of the dynamical system, we
study the perturbations in turn of the critical points. This drive
us to a general expression $T=M \delta$, where
\begin{equation} T=\left(
                     \begin{array}{c}
                        \frac{d \delta \rho_{de}}{d s} \\
                        \frac{d \delta \rho_{dm}}{d s} \\
                     \end{array}
                   \right), \qquad \delta =\left(
                     \begin{array}{c}
                       \delta \rho_{de} \\
                       \delta \rho_{dm} \\
                     \end{array}
                   \right),\label{as5}\end{equation}
and $M$ is the matrix (\ref{matrix}).


The corresponding eigenvalues are given by
\begin{equation}\label{as7}
  \lambda_{1, 2}=-\frac{3}{4}\left(4+3\omega_{de}+\omega_{dm} \pm  \kappa\right),
\end{equation}
which shows that the system is stable if and only if the condition
$\kappa<4+3\omega_{de}+\omega_{dm}$ is satisfied ($\lambda_1<0$).
The physically acceptable solution of this inequality can be written
as
\begin{equation}\label{as8}
  -1<\omega_{dm}<1, \quad \textrm{and}\quad
 -\frac{1}{2+\omega_{dm}}< \omega_{de}\leq \omega_{dm},
\end{equation}
which we have plotted in Fig. \ref{f1} as the shaded area in the $\omega_{de}$-$\omega_{dm}$ plane.


\section{The observational support}

\begin{figure}[!h]
  \begin{center}
    \includegraphics[width=80mm]{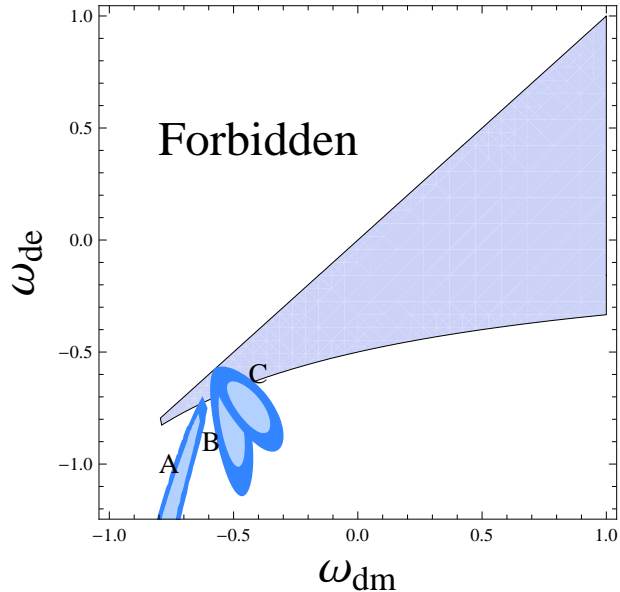}
  \end{center}
  \caption{Here we show the confidence contour for each case described in Table 1 together the forbidden
region in the free parameters obtained in the previous section. Note
that the three confidence contours are centered in the forbidden
region.}
  \label{f1}
\end{figure}

Using the analytical solution presented in the previous section we
can test it against the observations. In this section we use the
latest supernova Ia data -- the Union 2.1 set -- consisting in 580
SNIa points \cite{Suzuki:2011hu}. The SNIa data give the distance
modulus as a function of redshift $\mu_{obs}(z)$. Theoretically the
distance modulus is a function of the cosmology through the
luminosity distance
\begin{equation}\label{dlzf}
d_L(z)=(1+z)\frac{c}{H_0}\int_0^z \frac{dz'}{E(z')},
\end{equation}
valid for a flat universe with $E(z)=H(z)/H_0$. Explicitly the
theoretical value is computed by $\mu(z)=
5\log_{10}[d_L(z)/\texttt{Mpc}]+25$. We fit the SNIa with the
cosmological model by minimizing the $\chi^2$ value defined by
\begin{equation}
\chi^2=\sum_{i=1}^{580}\frac{[\mu(z_i)-\mu_{obs}(z_i)]^2}{\sigma_{\mu
i}^2},
\end{equation}
Assuming a flat geometry with today values $\Omega_{dm}=0.24$ and
$\Omega_{de}=0.76$ we obtain a best fit with $\chi^2_{min}=562.21$
and best fit values $\omega_{de}=-0.82$ and $\omega_{dm}=-0.51$ with
confidence contours at $68.27\%$ and $95.45\%$ shown in Figure
\ref{f1}. In this plot we also display the allowed region in
the parameter space for which the solution exist and is stable discussed in the last section.
Furthermore, to compare the performance of the model, we also consider the result in two extreme
cases: $\Omega_{dm}=0$, $\Omega_{de}=1$ and $\Omega_{dm}=1$,
$\Omega_{de}=0$, whose results are displayed in Table I, showing that although it is possible to find a set
of parameter that fit the SNIa data, as can be seen from the figure,
they all fall into a forbidden region even at one sigma. The
consequences of this are discussed in the next section.

\begin{table}[h!]\label{tab:table02}
\caption{The best fit values for the free parameters using the SNIa
data in a flat universe. We also show the $\chi^2_{min}$ of the fit
divided by the effective degrees of freedom. In Fig. \ref{f1} we
show the confidence contour for each case together the forbidden
region in the free parameters.}
\begin{center}
\begin{tabular}{cccccc}
Case & $\Omega_{dm}$ & $\Omega_{de}$ & $\chi^2_{min}/d.o.f.$ & $\omega_{dm}$ & $\omega_{de}$ \\
\hline
 A & 1    &  0   & 562.44/578 & -0.78 & -1.3 \\
 B & 0.24 &  0.76 & 562.21/578 & -0.51 & -0.83 \\
 C & 0    &  1   & 562.21/578 & -0.44 & -0.74 \\
\end{tabular}
\end{center}
\end{table}

\begin{figure}[!h]
  \begin{center}
    \includegraphics[width=130mm]{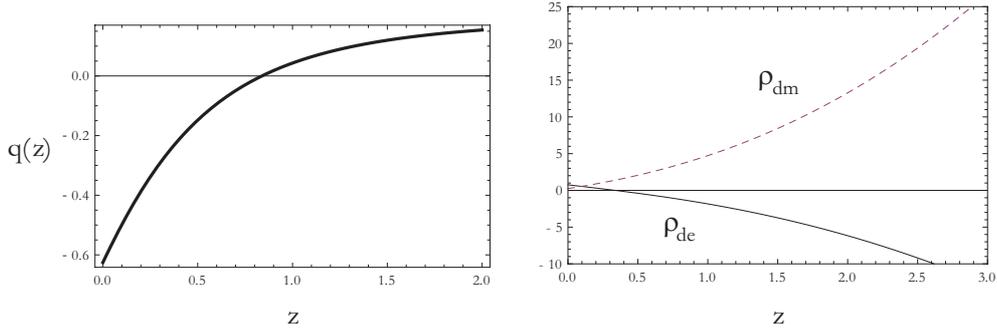}
  \end{center}
  \caption{Left Panel: The deceleration parameter reconstructed from the best fit $H(z)$ function.
  Right Panel: Here we plot the analytic solutions (\ref{as1}) of the energy
  densities, $\rho_{de}$ and $\rho_{dm}$ as a function of redshift, using
  the best fit values for the parameters found in the previous section. Notice
  how $\rho_{de}$ falls to negative values near $z\simeq 0.3$.}
  \label{f2}
\end{figure}

This means that we can fit the supernova data without using an
explicit cosmological constant, as was stressed in
\cite{Zhang:2010iz}. Actually, reconstructing the deceleration
parameter as a function of the redshift $q(z)$ (see Fig. \ref{f2})
we conclude that this model describes an accelerated expansion
without a cosmological constant.

\section{Violation of the weak energy condition}

However, there is a fatal failure in this model: there is no chance
to satisfied the weak energy condition. This can be viewed from the
explicit solution we have found (\ref{as1}). In fact, based on the
best fit values found in the previous section, we plot the energy
densities as a function of redshift in Figure \ref{f2}, where it is
clear that although both energy densities are positive at $z=0$, one
of them (in this case $\rho_{de}$) falls below zero near $z\simeq
0.3$.

In general, assuming initial values $\rho_{de}^{(0)}\geq 0$ and
$\rho_{dm}^{(0)}\geq 0$ in the regime where $\kappa$ is real, it is
clear that both factors $a$ and $b$ in (\ref{as1})
\textit{should} be negative simultaneously to get both energy densities 
positively defined. However, as a direct computation shows, if we have 
one of these parameters positive (negative), 
the other results to be negative (positive).  

\section{Summary}
In this paper we have reviewed the semi-holographic model presented by Zhang, Li and Noh \cite{Zhang:2010iz}.
From this, we have found the analytical solutions to the dynamical system generated by this model, Eqs. (\ref{as1}), without a cosmological constant.
Also, we have studied the stability of the system, which drive us to impose
constraints on the EoS parameters, $\omega_{dm}$ and $\omega_{de}$, Eqs. (\ref{cond1}) and (\ref{as8}). Based on this analysis, we determine the allowed region in parameter space to get stable solutions, which does not agree with those mentioned in \cite{Zhang:2010iz} (see eqs.
(23) and (24) in this reference). The analytical solution of the model was also tested against the latest available observational data from Supernovas Ia,
which is a set of $580$ points for the module distance. Assuming a flat geometry with today 
values $\Omega_{dm} = 0.24$ and $\Omega_{de} = 0.76$, we obtain a best
fit with $\chi^2_{min} = 562.21$ and best fit values $\omega_{de} = -0.82$ 
and $\omega_{dm} = -0.51$ with confidence contours at $68.27\%$ and 
$95.45\%$ shown in Figure (\ref{f1}). We have also considered two other cases whose results are displayed in Table I and Figure 1.
However, our analysis show that this model is not realistic because it does not satisfy the weak energy condition.

\begin{acknowledgments}
We wish to acknowledge useful conversations with S. del Campo. JM acknowledges ESO - Comit\'e Mixto,
VHC acknowledges financial support through DIUV
project No. 13/2009, and FONDECYT Grant N$^{\circ}$ 1110230.

\end{acknowledgments}


\begin{thebibliography}{99}

\bibitem{01}A. G. Riess et al., Astron. J. 116, 1009
(1998)[astro-ph/9805201 ]; S. J. Perlmutter et al., Astrophys. J.
517, 565(1999); A. G. Riess et al., Astrophys. J. 607, 665(2004).

\bibitem{02}M. Tegmark et al. [SDSS Collaboration], Phys. Rev. D 69, 103501
(2004); K. Abazajian et al. [SDSS Collaboration], Astron. J. 128,
502 (2004); K. Abazajian et al. [SDSS Collaboration], Astron. J.
129, 1755 (2005).

\bibitem{03} H. V. Peiris et al., Astrophys. J. Suppl. 148 (2003) 213 [astro-ph/0302225]; C. L.
Bennett et al., Astrophys. J. Suppl. 148  1 (2003); D. N. Spergel et
al., Astrophys. J. Suppl. 148  175 (2003).

\bibitem{isw}
S. Boughn and R. Chrittenden, Nature (London) \textbf{427}, 45
(2004); P. Vielva, E. Mart\'{\i}nez--Gonz\'{a}lez, and M. Tucci,
Mon. Not. R. Astron. Soc. \textbf{365}, 891 (2006).

\bibitem{Eisenstein:2005su}
  D.~J.~Eisenstein {\it et al.}  [SDSS Collaboration],
  Astrophys.\ J.\  {\bf 633}, 560 (2005)
  [astro-ph/0501171].

\bibitem{weakl}
C.R. Contaldi, H. Hoekstra, and A. Lewis, Phys. Rev. Lett.
\textbf{90}, 221303 (2003).

\bibitem{1} S. Weinberg, Rev. Mod. Phys. 61, 1 (1989);
 E.J. Copeland, M. Sami, S. Tsujikawa, Int. J. Mod. Phys. D 15, 1753 (2006).

\bibitem{2}C. Wetterich, Nucl. Phys. B 302, 668 (1988); B. Ratra, J. Peebles,
Phys. Rev. D 37, 321 (1988).

\bibitem{3}T. Chiba, T. Okabe, M. Yamaguchi, Phys. Rev. D 62, 023511 (2000);
C. Armend´ariz-Pic´on, V. Mukhanov, P.J. Steinhardt, Phys. Rev.
Lett. 85, 4438 (2000).

\bibitem{4} A. Sen, J. High Energy Phys. 10, 008 (1999); E.A. Bergshoeff, M.
de Roo, T.C. de Wit, E. Eyras, S. Panda, J. High Energy Phys. 05,
009 (2000).

\bibitem{interaction} C. Wetterich, Nucl. Phys. B \textbf{302},
668 (1988); {\em ibid.} Astron. Astrophys. \textbf{301}, 321 (1995).

\bibitem{5}A. Kamenshchik, U. Moschella, V. Pasquier, Phys. Lett. B 511, 265
(2001).

\bibitem{bv} W.S. Hip\'{o}lito-Ricaldi, H.E.S. Velten and W. Zimdahl,
Phys. Rev. D  \textbf{82}, 063507 (2010).

\bibitem{7} G. 't Hooft, gr-qc/9310026; L. Susskind, J. Math. Phys. 36, 6377
(1995).

\bibitem{FS98}W.~Fischler and L.~Susskind, arXiv:hep-th/9806039 (1998).

\bibitem{cohen}
A. G. Cohen, D.B. Kaplan and A.E. Nelson, Phys. Rev. Lett.
\textbf{82}, 4971 (1999).

\bibitem{li}
M. Li, Phys. Lett. B \textbf{603}, 1 (2004).

\bibitem{021}C. Gao, F. Q. Wu, X. Chen and Y. G. Shen, Phys. Rev. D
{\bf 79}, 043511 (2009).

\bibitem{022} C. J. Feng, Phys. Lett. B {\bf 670}, 231 (2008);
~L. N. Granda and A. Oliveros, Phys. Lett. B {\bf 669}, 275 (2008).

\bibitem{Zhang:2010iz}
  H.~Zhang, X.~-Z.~Li and H.~Noh,
  Phys.\ Lett.\ B {\bf 694}, 177 (2010)
  [arXiv:1010.1362 [gr-qc]].

\bibitem{Li:2012vw}
  H.~Li, H.~Zhang and Y.~Zhang,
  arXiv:1212.2360 [astro-ph.CO].

\bibitem{cai}
  R.~G.~Cai and S.~P.~Kim,
  JHEP {\bf 0502}, 050 (2005)
  ,[hep-th/0501055].

 \bibitem{bak}
 D. Bak and S. J. Rey, Class. Quant. Grav. 17, L83 (2000)
 [arXiv:hep-th/9902173].

\bibitem[{{Suzuki et al.}(2011)}]{Suzuki:2011hu}
Suzuki N., Rubin D., Lidman C., Aldering G., Amanullah R., Barbary
K., Barrientos L.F. and Botyanszki J. {\it et al.}, 2011,
  ApJ 746, 85



\end{thebibliography}
\end{document}